\PassOptionsToPackage{clashing-option}{package-name}
\PassOptionsToPackage{mainaux}{rerunfilecheck}
\documentclass[dvipsnames,format=sigconf,authorversion, nonacm, preprint]{acmart}
\renewcommand\footnotetextcopyrightpermission[1]{} % Removes footnote with conference info
\setcopyright{none} % Suppresses copyright block

\usepackage{algorithm}
\usepackage{algpseudocode}
\usepackage{multirow}
\usepackage{booktabs}
\usepackage{graphicx}
\usepackage{colortbl}
\usepackage[table]{xcolor} % For \cellcolor
\usepackage{pgf}           % For math calculations
\usepackage{makecell}      % CRITICAL: Allows \\ inside table cells
\DeclareUnicodeCharacter{2248}{\ensuremath{\approx}}
\definecolor{tie}{RGB}{255,235,190} % light orange tie
\newcommand{\gradcell}[4]{%
  \ifnum#2=#3%
    \cellcolor{tie}\makecell{#1/#2/#3}%
  \else%
    \pgfmathtruncatemacro{\diff}{abs(#2-#3)}%
    \pgfmathtruncatemacro{\shade}{min(100, round(100*\diff/#4))}%
    \ifnum#2>#3%
      \edef\temp{\noexpand\cellcolor{red!\shade}}\temp%
    \else%
      \edef\temp{\noexpand\cellcolor{green!\shade}}\temp%
    \fi%
    \makecell{#1/#2/#3}%
  \fi%
}
\begin{document}

% \title{Quantum-Assisted Discovery of Diverse Modes in Black-Box Landscapes via Surrogate-Based Sampling}
\title{Surrogate-Guided Quantum Discovery in Black-Box Landscapes with Latent-Quadratic Interaction Embedding Transformers}
\author{Saisubramaniam Gopalakrishnan, Dagnachew Birru}
\affiliation{%
  \institution{Philabs, Quantiphi}
  \country{India, USA}
}
\email{{gopalakrishnan.saisubramaniam, dagnachew.birru}@quantiphi.com}

% \renewcommand{\shortauthors}{Gopalakrish et al.}

\begin{abstract}
Discovering configurations that are both high-utility and structurally diverse under expensive black-box evaluation and strict query budgets remains a central challenge in data-driven discovery. Many classical optimizers concentrate on dominant modes, while quality-diversity methods require large evaluation budgets to populate high-dimensional archives. Quantum Approximate Optimization Algorithm (QAOA) provides distributional sampling but requires an explicit problem Hamiltonian, which is unavailable in black-box settings. Practical quantum circuits favor quadratic Hamiltonians since higher-order interaction terms are costly to realize. Learned quadratic surrogates such as Factorization Machines (FM) have been used as proxies, but are limited to pairwise structure.
We extend this surrogate-to-Hamiltonian approach by introducing a Latent-Quadratic  Interaction Embedding Transformer (QET) that models higher-order variable dependencies via self-attention and projects them into a valid Positive Semi-Definite quadratic form compatible with QAOA. This enables diversity-oriented quantum sampling from learned energy landscapes while capturing interaction structure beyond pairwise terms.
We evaluate on risk discovery for enterprise document processing systems against diverse classical optimizers. Quantum-guided samplers achieve competitive utility while consistently improving structural diversity and exclusive discovery. FM surrogates provide stronger early coverage, whereas QET yields higher-fidelity surrogate landscapes and better extreme-case discovery. QET–QAOA methods recover roughly twice as many structurally tail-risk outliers as most classical baselines and identify an exclusive non-overlapping fraction (≈4–5\%) of high-utility configurations not found by competing methods. 
These results highlight Latent-Quadratic Interaction embedding with self-attention as an effective mechanism for learning higher-order interaction structure and projecting it into quadratic surrogate Hamiltonians for quantum-assisted black-box discovery.
\end{abstract}

% <concept_id>10010147.10010257</concept_id>
% <concept_desc>Computing methodologies~Machine learning</concept_desc>
% <concept_id>10010520.10010521.10010542.10010550</concept_id>
% <concept_desc>Computer systems organization~Quantum computing</concept_desc>
% <concept_id>10010147.10010257.10010293.10010319</concept_id>
% <concept_desc>Computing methodologies~Learning latent representations</concept_desc>
% <concept_significance>300</concept_significance>
% <concept_id>10010147.10010257.10010282.10011304</concept_id>
% <concept_desc>Computing methodologies~Active learning settings</concept_desc>
% <concept_significance>300</concept_significance>
% \ccsdesc[500]{Computing methodologies~Machine learning}
% \ccsdesc[500]{Computer systems organization~Quantum computing}
% \ccsdesc[300]{Computing methodologies~Learning latent representations}
% \ccsdesc[300]{Computing methodologies~Active learning settings}



\fancypagestyle{firstpage}{
  \fancyhf{}
  \renewcommand{\footrulewidth}{0pt}    % Add footer line (0.4pt is standard)
  \fancyfoot[L]{\\\textit{Preprint. Under Review.}}
}
\maketitle
\thispagestyle{firstpage} % Apply the style here
\section{Introduction}
Discovering configurations that are simultaneously high-utility and structurally diverse in complex black-box landscapes is a fundamental challenge in data-driven science and engineering. This need for high-coverage discovery spans domains as diverse as {\textit{de novo} molecular design} \cite{gomez2018automatic}, {financial stress testing} \cite{glasserman2004monte}, and the {reliability assurance of Intelligent Document Processing (IDP)} systems \cite{cui2021document}, where the objective is rarely a single global optimum but rather a diverse portfolio of high-performing solutions. This setting is especially critical in validation workflows, where uncovering multiple distinct failure modes is essential for robust assurance. However, such problems are typically governed by expensive black-box oracles and strict query budgets, rendering exhaustive search infeasible.

Classical approaches rely primarily on population-based heuristics, yet they exhibit well-known limitations under tight evaluation budgets. Evolutionary and swarm-based methods effectively exploit local information but often suffer from mode collapse, converging to narrow clusters around dominant basins of attraction \cite{crepinsek2013exploration}. Quality-Diversity (QD) methods such as MAP-Elites explicitly enforce behavioral diversity through archive-based illumination \cite{mouret2015illuminating, pugh2016qd}, but their archive-filling dynamics are frequently sample-inefficient in high-dimensional spaces, often requiring orders of magnitude more evaluations than are available in industrial testing regimes.
Bayesian Optimization (BO) and reinforcement learning (RL) methods are query-efficient for sequential decision-making, but are typically formulated to optimize scalar objectives or expected return, and require additional novelty or entropy regularization to support diverse solution set discovery \cite{shahriari2015taking, maus2023discovering}.
As a result, there remains a gap between query-efficient optimization and principled discovery of structurally diverse high-utility configurations.

Quantum optimization offers a complementary mechanism for maintaining solution diversity through probabilistic sampling over an energy landscape. However, algorithms such as the Quantum Approximate Optimization Algorithm (QAOA) \cite{farhi2014quantum} require an explicit problem Hamiltonian, which is not directly accessible in black-box settings. Learned quadratic surrogates such as Factorization Machines (FM) \cite{rendle2010factorization} partially address this gap by modeling pairwise interactions, and have recently been coupled with quantum optimizers in black-box design pipelines \cite{tamura2023blackbox}. Nevertheless, such models are limited in their ability to represent higher-order dependencies and are typically applied to optimization rather than diverse discovery.

To address these limitations, we propose the \textit{Latent-Quadratic Interaction Embedding Transformer (QET)}, a surrogate-guided quantum discovery framework that learns a quadratic proxy Hamiltonian directly from observational data. QET employs a Transformer encoder with self-attention to model higher-order dependencies among decision variables, and projects the learned interactions into a valid positive semi-definite (PSD) quadratic form compatible with Ising/QUBO representations. This surrogate Hamiltonian is then used within QAOA at low alternating-operator depth, enabling the quantum circuit to function as a structured diversity-oriented sampler rather than a traditional ground-state optimizer. By adjusting circuit parameters and mixing operators, the sampler concentrates probability mass around multiple low-energy basins while preserving multi-modal coverage.
We empirically evaluate the framework on high-dimensional reliability testing of enterprise Intelligent Document Processing (IDP) systems, which serves as a representative High-dimensional Expensive Problem (HEP) \cite{Zhou2024Survey} setting. The search space is combinatorial, the oracle is computationally costly, and the resulting risk landscape is fragmented and non-differentiable. 
% The underlying structure of assembling discrete components to minimize a black-box score makes this benchmark representative of a broader class of scientific discovery problems.
The underlying structure of assembling discrete document components to optimize a black-box risk score makes this benchmark representative of a broader class of other scientific discovery problems.

Our contributions are as follows:

\begin{itemize}
    \item \textbf{Latent Proxy Hamiltonian Learning:} We introduce a novel surrogate architecture \textit{Latent-Quadratic Interaction Embedding Transformer (QET)} that bridges deep learning and quantum. Unlike standard FM limited to fixed pairwise interactions, QET employs a self-attention mechanism to capture complex, higher-order dependencies in the black-box objective. We introduce a projection layer that maps these latent embeddings onto a valid, {Positive Semi-Definite (PSD) quadratic form}, effectively constructing a realizable Ising Hamiltonian from observational data.

    \item \textbf{Quantum Diversity Sampling:} We reframe the role of QAOA from ground state optimization to diversity sampling. We demonstrate that by driving the ansatz with the learned QET proxy, the induced wavefunction leverages quantum superposition to simultaneously populate distinct low-energy basins, effectively bypassing the mode collapse typical of greedy classical heuristics.

    \item \textbf{Structural Exclusivity \& Outlier Quantification:} We validate the framework on a document processing risk discovery benchmark against 10 state-of-the-art solvers and their variants. We demonstrate that QET achieves superior predictive fidelity ($R^2 \approx 0.84$), through set-theoretic analysis it isolates a distinct utility subspace (4--5\% of yield exclusive to the union of other baselines), and alongside FM-surrogate, it ranks amongst the top in discovering tail 1\% edge cases, making it an important surrogate mechanism for robust quantum-based discovery.
\end{itemize}

\begin{figure*}[t]
  \centering
  \includegraphics[width=1.0\linewidth]{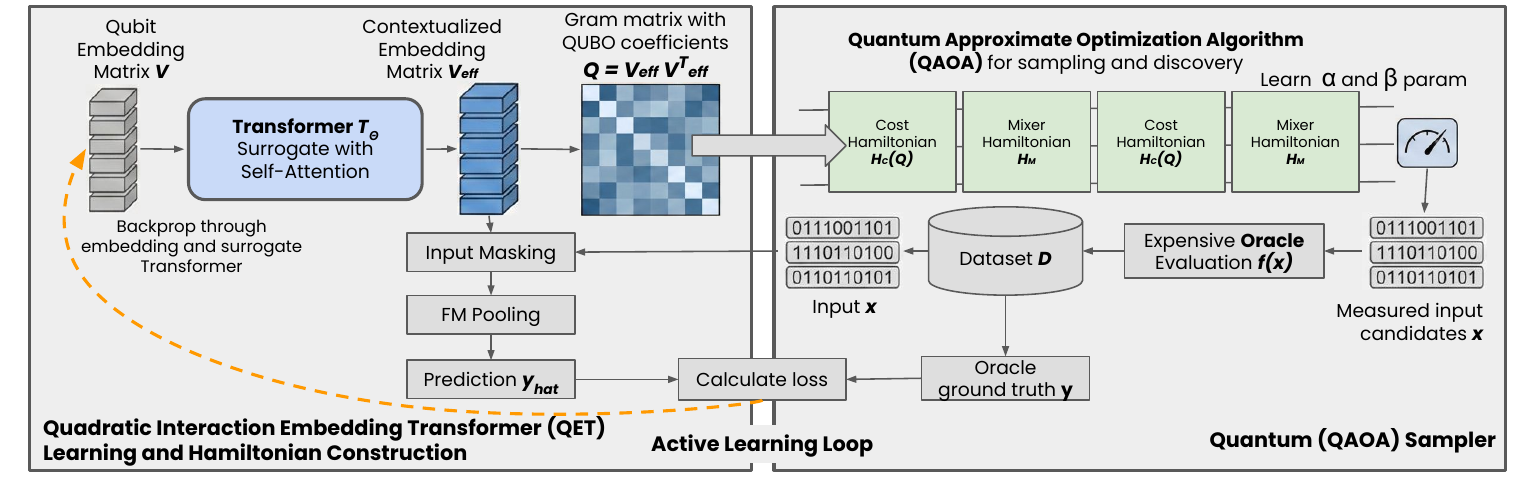}
\caption{
The QET-QAOA Discovery Framework. (A) Latent-Quadratic Interaction Embedding Transformer: Contextualizes latent embeddings $V$ via self-attention in Transformer $\mathcal{T}_\theta$ to capture higher-order dependencies, predicting $\hat{y}$ via Factorization Machine pooling.
(B) QUBO and Hamiltonian Projection: The learned contextual embeddings $V_{\text{eff}}$ forms a Gram matrix $Q = V_{\text{eff}} V_{\text{eff}}^T$,
projecting the learned topology onto a valid PSD matrix that defines the QUBO coefficients, which are mapped to an equivalent Ising Hamiltonian $H_C$.
(C) QAOA Sampling: This proxy $H_C$ drives a QAOA circuit to sample diverse high-quality configurations, which are evaluated by the Oracle and fed back to iteratively refine the surrogate.
}
\label{fig:main_fig}
\end{figure*}

% \vspace{-0.3cm}
\section{Related Work}
We situate our contributions at the intersection of budget-constrained black-box optimization, diversity-driven search, and quantum-assisted machine learning. This section contrasts our framework with existing classical heuristics, classical surrogate and policy learning approaches, and prior work in quantum for optimization.

\subsection{Optimizers and Diversity-Driven Discovery}
The challenge of identifying sets of solutions that are both high-performing and structurally distinct (termed Quality-Diversity (QD) or Illumination) is central to complex systems discovery. Standard population-based heuristics, such as Genetic Algorithms (GA) \cite{holland1975adaptation} and Particle Swarm Optimization (PSO) \cite{kennedy1995pso}, prioritize gradient-free exploitation but frequently suffer from \emph{mode collapse}. These methods often converge to a single basin of attraction, leaving structurally distinct high-utility regions unexplored \cite{crepinsek2013exploration}.
To mitigate this, QD algorithms like MAP-Elites \cite{mouret2015illuminating} and Novelty Search with Local Competition (NSLC) \cite{lehman2011evolving} explicitly mandate diversity by maintaining archives of solutions indexed by behavioral descriptors. While QD methods excel at coverage, they are notoriously \emph{sample inefficient}, typically requiring hundreds of thousands of evaluations ($>10^5$) to illuminate high-dimensional feature spaces \cite{pugh2016qd}.
This renders them impractical for industrial settings governed by expensive black-box oracles. Our framework bridges this gap: by using  QET to learn the \textit{latent topology} of the landscape, we construct a surrogate that enables the solver to mimic diversity-seeking behavior of QD methods, but with the sample efficiency of optimizers.

\subsection{Surrogate-Based and Policy Optimization}
For expensive black-box functions, Bayesian Optimization (BO) \cite{snoek2012practical} remains the standard for sample efficiency. By utilizing probabilistic surrogates such as Gaussian Processes (GP) or Tree-structured Parzen Estimators (TPE) \cite{bergstra2011tpe}, BO guides the search toward promising regions. However, standard kernels (e.g., Matérn or Hamming) often struggle to capture complex, non-local variable dependencies in high-dimensional combinatorial spaces.
To address this, Factorization Machines (FM) \cite{rendle2010factorization} and their neural variants, such as DeepFM \cite{guo2017deepfm} and Attentional Factorization Machines (AFM) \cite{xiao2017attentional}, have emerged as powerful surrogates. Unlike linear models, FMs explicitly model second-order interactions 
% ($\langle \mathbf{v}_i, \mathbf{v}_j \rangle x_i x_j$), 
and have been successfully applied to material design and discovery \cite{nature2025higher, kitai2020designing}.

Parallel to surrogate modeling, Deep Reinforcement Learning (DRL) has been adapted for combinatorial optimization, treating the search as a sequential decision process. Policy gradient methods, specifically {REINFORCE} \cite{williams1992simple} and {Proximal Policy Optimization (PPO)} \cite{schulman2017proximal}, have shown promise in learning constructive heuristics. However, these methods are typically data-hungry and require exploration to stabilize the policy gradient in sparse-reward environments, making them less suitable for strictly budget-constrained discovery compared to surrogate-based approaches.

\subsection{Quantum-Assisted Learning and Sampling}
The integration of quantum solvers with classical surrogates, e.g. FMQA \cite{kitai2020designing}, has demonstrated that learned FM weights can be directly mapped to Quadratic Unconstrained Binary Optimization (QUBO) formulations for Quantum Annealers (QA) \cite{kadowaki1998quantum}. However, existing FMQA literature focuses strictly on scalar minimization to collapse the system to a single global optimum.
Our work diverges by repurposing this pipeline for diversity sampling. We leverage the variational gate-based QAOA \cite{farhi2014quantum} algorithm to sample from a multi-modal wavefunction, effectively addressing the {budgeted discovery} \cite{guha2007approximation} problem where finding a diverse portfolio of candidates is superior to point-convergence.
Furthermore, while standard QAOA relies on unconstrained transverse field mixing, our framework aligns with the {Quantum Alternating Operator Ansatz (QAOA+)} \cite{hadfield2019quantum} by utilizing interaction-aware mixing from the leared QUBO to preserve domain constraints, ensuring that the quantum sampler explores only the feasible subspace defined by the domain.

% \vspace{-0.3cm}
\subsection{Generative Circuit Synthesis vs. Higher-Order Latent Hamiltonian Learning}
% Recent advances have focused on neural circuit synthesis and parameter transfer, where models like {QAOA-GPT} \cite{arxiv2025qaoagpt}, Graph Attention Networks (GAN) \cite{arxiv2025transfer}, or Quantum Transformers \cite{surveyquantumtransformers2025} are trained to predict optimal QAOA parameters ($\gamma, \beta$) for a given problem instance.
% Similarly, dynamic ansatz construction methods such as ADAPT-QAOA \cite{zhu2022adaptive} iteratively grow the quantum circuit by selecting operators from a pool based on energy gradients, achieving high accuracy on NISQ devices.
Recent work explores neural circuit synthesis and parameter transfer, where models such as QAOA-GPT \cite{arxiv2025qaoagpt}, Graph Attention Networks \cite{arxiv2025transfer}, and quantum transformers \cite{surveyquantumtransformers2025} predict instance-specific QAOA parameters $(\gamma,\beta)$. Complementarily, dynamic ansatz methods like ADAPT-QAOA \cite{zhu2022adaptive} grow circuits by selecting operators via energy gradients, achieving strong performance on NISQ devices.
These approaches assume that the problem graph (the Hamiltonian $H_C$) is \textit{known a priori} (e.g., MaxCut on a provided graph). They learn the mapping $\mathcal{G} \to \text{Circuit}$.
In contrast, our work addresses and operates under a \textit{black-box setting} where the underlying interaction graph is unknown and must be inferred from limited data. Our QET architecture performs \textit{Quadratic Structure Learning} (Data $\to \mathcal{G}$), effectively serving as the necessary precursor step that makes quantum optimization applicable to black-box tasks.

Finally, while recent works have explored higher-order (HUBO) formulations for portfolio optimization \cite{huboportfolio2025} and proposed native multi-body pulses via Digital-Analog Quantum Computing (DAQC) \cite{parra2020digital}, we explicitly constrain our architecture to the Ising model. Implementing native higher-order gates (e.g., $ZZZ$) requires costly circuit decompositions that limit scalability on Near-Term (NISQ) devices. By using a Transformer to capture higher-order dependencies in the \textit{embedding space} and projecting them onto a quadratic form,  QET captures the effective spectral influence of complex interactions without incurring the hardware overhead of a full HUBO implementation.

% \subsection{Transferability and Generalized Quantum Learning}
% A key limitation of standard QAOA is the need to re-optimize parameters from scratch for every instance. Recent "Quantum Transformer" approaches \cite{surveyquantumtransformers2025} suggest that attention mechanisms can learn generalized heuristics. By learning a global latent representation $V$ rather than instance-specific scalars, our QET architecture aligns with this paradigm, potentially enabling transfer learning across different black-box functions of similar classes.

\section{Methodology}
\label{section:methodology}
Figure \ref{fig:main_fig} illustrates the end-to-end QET-QAOA workflow, which iterates through three phases: (A) learning a latent structural embedding of Latent-Quadratic Interactions from data, (B) projecting these embeddings into a valid Cost Hamiltonian to QAOA, and (C) utilizing the induced quantum energy landscape for diversity sampling.

\subsection{Latent-Quadratic Interaction Embedding}\subsubsection{Surrogate Modeling:} To enable the use of QUBO/Ising Hamiltonian inside QAOA, we first need to approximate the black-box risk function $f(\mathbf{x})$ into a quadratic model. The Latent-Quadratic Interaction Embedding Transformer (QET) model is built upon a global learnable latent matrix $V \in \mathbb{R}^{d \times k}$ comprising individual Latent-Quadratic interactive information as a vector or embedding.

\textbf{1. Embedding Contextualization:} To capture higher-order dependencies, the matrix $V$ is processed by a Transformer encoder $\mathcal{T}_\theta$. The self-attention mechanism contextualizes the latent vectors based on their global relationships, outputting a refined matrix $V_{\text{eff}} = \mathcal{T}_\theta(V)$.
% , where each row $\mathbf{v}'_i$ represents the $i$-th variable enriched with system-wide context.

\textbf{2. Masked Pooling and Prediction:} The input bitstring $\mathbf{x}$ acts as a \textit{masking operator}. 
% We define masked latent vectors $\mathbf{v}''_i = \mathbf{v}'_i x_i$.
The scalar risk score $\hat{y}$ is then computed using the efficient linear-time factorization of the pairwise interactions using a second-order Factorization Machine (FM) pooling layer as given in \cite{rendle2010factorization}:
\begin{equation}
    \hat{y}(\mathbf{x}) = w_0 + \sum\limits_{i=1}^d w_i x_i + \frac{1}{2} \sum\limits_{f=1}^k \left[ \left( \sum\limits_{i=1}^d v_{i,f} x_i \right)^2 - \sum\limits_{i=1}^d (v_{i,f} x_i)^2 \right]
\end{equation}
Here, the inner term computes the interactions by subtracting the sum of squared elements from the square of the sum of elements, allowing the model to capture all $O(d^2)$ pairwise interactions in just $O(dk)$ time.
Here, $f$ represents the index over the embedding dimensions, ranging from $1$ to $k$. The term $v_{i,f}$ denotes the $f$-th component of the masked latent vector for variable $i$. 
This formulation is fully differentiable. During the learning phase, the error signal from the prediction $\hat{y}$ backpropagates through the pooling layer and the Transformer encoder $\mathcal{T}_\theta$, updating both the attention weights and the initial latent embeddings $V$. The entire architecture is trained end-to-end to minimize the Mean Squared Error (MSE) $\mathcal{L} = \frac{1}{N} \sum (y_{\text{true}} - \hat{y})^2$, on the history of observed configurations.

\subsubsection{Hamiltonian Extraction}Post-training, we extract the effective latent matrix $V_{\text{eff}}$. The QUBO interaction matrix $Q$ is constructed as the Gram matrix of these refined vectors, which projects the learned problem surface onto a quadratic form:\begin{equation}Q = V_{\text{eff}} V_{\text{eff}}^T\end{equation}
% We map $Q$ to an Ising Hamiltonian $H_C = \sum J_{ij} Z_i Z_j + \sum h_i Z_i$ using the standard transformation $x_i \to (1-Z_i)/2$.

% \subsubsection{Quantum-Assisted Sampling via QAOA}We utilize the Quantum Approximate Optimization Algorithm (QAOA) \cite{farhi2014quantum} as a \textit{diversity inducing sampler} instead of a conventional solver by using a low circuit depth $p=2$. The learned surrogate $H_C$ serves as the phase-separation operator. We optimize the variational parameters $(\boldsymbol{\gamma}, \boldsymbol{\beta})$ to minimize the expectation value $\langle H_C \rangle$. The resulting quantum state $|\psi(\boldsymbol{\gamma}, \boldsymbol{\beta})\rangle$ is a superposition concentrated over low-energy (or negative of high energy viz.) basins. Repeated measurement of this state yields a batch of candidates that are both high-quality (low risk) and diverse. 
% % These candidates are evaluated by the oracle and added to the dataset  $\{\mathbf{x}_{\text{new}}, y_{new}\}$ to close the discovery loop.
% These candidates are evaluated by the oracle to obtain their true risk scores. The training dataset $\mathcal{D}$ is then updated via $\mathcal{D} \leftarrow \mathcal{D} \cup \{(\mathbf{x}_i, y_i)\}_{i=1}^B$, where $B$ is the budget size, closing the iterative discovery loop.
\subsection{Quantum-Assisted Sampling via QAOA}
% To generate diverse candidate solutions, 
We utilize the Quantum Approximate Optimization Algorithm (QAOA) \cite{farhi2014quantum} as a \textit{diversity inducing sampler} instead of a conventional solver by using a low circuit depth $p=2$ and
% we utilize the Quantum Approximate Optimization Algorithm (QAOA) \cite{farhi2014quantum}. Unlike standard optimization where the goal is a single ground state, we employ QAOA as a \textit{generative sampler}, 
 preparing a wavefunction $|\psi(\boldsymbol{\gamma}, \boldsymbol{\beta})\rangle$ concentrated on low-energy basins.\subsubsection{Cost Hamiltonian Construction}First, we map the learned QUBO matrix $Q$ to an Ising Hamiltonian $H_C$ using the transformation $x_i \to (1-Z_i)/2$. This defines the problem landscape:\begin{equation}H_C = \sum_{i<j} J_{ij} Z_i Z_j + \sum_i h_i Z_i\end{equation} $J_{ij}$ and $h_i$ are derived directly from the projected Gram matrix $Q$.
 \subsubsection{Topology-Aware Mixing Strategies} An additional component of our framework is the choice of the Mixer Hamiltonian $H_M$, which governs how the quantum state explores the search space. To leverage the structural information captured by QET, we investigate two distinct mixing strategies:\begin{enumerate}\item \textbf{Standard Transverse Mixer (QAOA):} The conventional approach using independent bit-flip operators, $H_M = \sum_i X_i$. This promotes global, unconstrained exploration but ignores variable dependencies.\item \textbf{Correlated Mixer (QAOA-Corr):} We propose a structure-aware mixer aligned with the QAOA+ framework \cite{hadfield2019quantum}. We identify the set of strongest pairwise dependencies $\mathcal{E}_{\text{top}}$ from the learned matrix $Q$ (i.e., pairs $(i,j)$ with the largest $|Q_{ij}|$). The mixer is augmented with correlated flip terms:
\begin{equation}
    H_M^{\text{corr}} = \sum_i X_i + \lambda \sum_{(i,j) \in \mathcal{E}_{\text{top}}} X_i X_j
\end{equation}
\end{enumerate}
By including the $X_i X_j$ terms, this operator induces \textit{correlated bit flips}, allowing the sampler to tunnel more effectively through the rugged landscape defined by strong variable interactions.
\begin{algorithm}[t]
\caption{QET-QAOA Surrogate-Guided Discovery}
\label{alg:QET_qaoa}
\begin{algorithmic}[1]
\Require Budget $B$, Oracle $f(\cdot)$, QET model $\mathcal{M}_\theta$, QAOA with $H_C$ Cost and $H_M$ Mixer Hamiltonian
\State Initialize dataset $\mathcal{D} \leftarrow$ Random Design of Experiments
\For{$t = |\mathcal{D}|$ to $B$}
    \State \textbf{1. Latent-Quadratic Interaction Embedding Learning:} Train QET $\mathcal{M}_\theta$ on $\mathcal{D}$ through FM projection to refine embeddings $V$
    \State \textbf{2. Hamiltonian Projection:} Compute Gram matrix $Q \leftarrow V V^T$ to form PSD proxy Hamiltonian $H_C$
    \State \textbf{3. Quantum Parameter Learning:} Optimize QAOA ansatz parameters $(\boldsymbol{\gamma}, \boldsymbol{\beta})$ to minimize $\langle H_C \rangle$ and $\langle H_M \rangle$
    \State \quad \textbf{4. Sampling:} Measure the wavefunction $|\psi(\boldsymbol{\gamma}, \boldsymbol{\beta})\rangle$ to sample diverse batch $\{\mathbf{x}'_1, \dots, \mathbf{x}'_k\}$
    \State \textbf{5. Oracle Evaluation:} Query $y'_i = f(\mathbf{x}'_i)$ for top-k candidates in batch
    \State \textbf{6. Active Update:} $\mathcal{D} \leftarrow \mathcal{D} \cup \{(\mathbf{x}'_i, y'_i)\}$
\EndFor
\State \Return Discovery Archive $\mathcal{D}$
\end{algorithmic}
\end{algorithm}
\subsubsection{Execution and Sampling}The circuit evolves the initial state $|+\rangle^{\otimes d}$ by alternating applications of the cost and mixer unitaries $p$ times:\begin{equation}|\psi(\boldsymbol{\gamma}, \boldsymbol{\beta})\rangle = \prod_{l=1}^p e^{-i \beta_l H_M} e^{-i \gamma_l H_C} |+\rangle^{\otimes d}\end{equation}Repeated measurement of the final state yields a batch of candidates that are evaluated by the oracle to obtain their true risk scores. The training dataset $\mathcal{D}$ is then updated via $\mathcal{D} \leftarrow \mathcal{D} \cup \{(\mathbf{x}_i, y_i)\}_{i=1}^B$, where $B$ is the budget size, closing the iterative discovery loop. 
% Algorithm \ref{alg:QET_qaoa} guides through the stages of the surrogate-guided discovery.

\subsection{Theoretical Properties}\subsubsection{Spectral Validity (PSD Constraint)}The parameterization $Q = V_{\text{eff}} V_{\text{eff}}^T$ enforces a rigid geometric constraint whereby the interaction matrix is guaranteed to be Positive Semi-Definite (PSD). For any vector $\mathbf{z}$, the energy $\mathbf{z}^T Q \mathbf{z} = ||V_{\text{eff}}^T \mathbf{z}||^2 \geq 0$. This ensures a smooth, convex-like energy surface, preventing the \textit{shattered} landscapes \cite{scholkopf2002learning} (indefinite matrices with many saddle points) that often plague unconstrained polynomial regression.
\subsubsection{Implicit Projection of Higher-Order Terms}While the true function $f(\mathbf{x})$ contains higher-order interactions ($x_i x_j x_k$),  QET performs a \textit{learned variational projection}. 
Formally, minimizing the MSE objective is equivalent to projecting the true higher-order risk function onto the manifold of Ising models, finding the closest pairwise approximation supported by the data \cite{mehta2019highbias}.
By training QET to minimize MSE, the model learns a latent configuration $V_{\text{eff}}$ that best approximates the marginal contribution of these higher-order terms. Consequently, the pairwise term $Q_{ij} = \langle \mathbf{v}'_i, \mathbf{v}'_j \rangle$ encodes the effective conditional influence of third-party variables, projecting the complex risk surface onto an optimal Ising basis without requiring auxiliary qubits \cite{kempe2006complexity}.

\subsection{Quantifying Discovery: Set-Theoretic Exclusivity}
\label{set_theoric_exclusivity}
To formally evaluate the \textit{marginal contribution} of the quantum sampler, we introduce the Exclusive Yield metric. Let $S_{\text{Q}}$ be the set of high-utility solutions found by our framework, and $S_{\text{Classical}}$ be the union of solutions found by all other classical baselines. The \emph{Exclusive Yield} ($\eta_{\text{ex}}$) is defined as:\begin{equation}\eta_{\text{ex}}(Q) = \frac{|S_{\text{Q}} \setminus S_{\text{Classical}}|}{|S_{\text{Q}} \cup S_{\text{Classical}}|}\end{equation}This metric rigorously quantifies the proportion of the high-utility landscape that is accessible \textit{solely} via the quantum-assisted approach.

\section{Experimental Evaluation}
\label{section:experiments}
We evaluate the proposed QET-QAOA framework against a comprehensive suite of 10 state-of-the-art baselines with variants. The primary objective is to validate how well quantum methods and the learned QET surrogate in particular effectively capture the complex problem landscape, enabling samplers to recover high-utility configurations accessible to standard heuristics. Our analysis is guided by two refined research questions: 
% \begin{itemize} \item \textbf{RQ1 (Comparative Discovery \& Efficacy):} How do Quantum and the two surrogates FM and QET compare to state-of-the-art classical Evolutionary, Bayesian, and RL baselines in discovering high-risk configurations under strict budgets? We evaluate performance across three dimensions: \emph{Exploitation-Exploration tradeoff} (Max and $\mu / \sigma$ risk scores), \emph{Topological Exclusivity} (Exclusive yield not found in other methods), and \emph{Structural Deviance and Landscape Fidelity} (Outlier discovery and surrogate fidelity).
% \item \textbf{RQ2 (Surrogate Ablation):} Does the {Latent-Quadratic Interaction Embedding Transformer (QET)} provide a measurable advantage over standard Factorization Machines (FM)?
% \end{itemize}
\textbf{RQ1 (Comparative Discovery):} How do quantum-guided methods using FM and QET surrogates compare against state-of-the-art evolutionary, Bayesian, and reinforcement learning baselines in discovering high-risk configurations under strict evaluation budgets, when all methods operate under identical oracle constraints? \textbf{RQ2 (Surrogate Ablation):} Does the Latent-Quadratic Interaction Embedding Transformer (QET) provide a measurable advantage over standard Factorization Machines (FM) in a controlled ablation study where the surrogate model is the only varying component?

\subsection{Case Study: Discoverying Risky Documents in IDP Pipeline Evaluation}
To simulate a realistic, high-dimensional black-box optimization task, we utilize an Enterprise Intelligent Document Processing (IDP) pipeline described in \cite{gopalakrishnan2026search}. 
% (Figure~\ref{fig:pipeline}) 
The optimization goal is \emph{Risk Document Discovery}: finding valid binary configurations $\mathbf{x} \in \{0,1\}^d$ that generate synthetic documents maximizing the failure rate of downstream OCR and extraction models. Each configuration specifies a structured document blueprint via binary-encoded structural parameters such as table density, noise artifacts, pagination behavior, and layout variations, which we refer to as \emph{risk features}.
% \begin{figure}[t]
%   \centering
%   \includegraphics[width=0.95, width=\linewidth]{figures/risk_discovery_figure.pdf}
%   \caption{Schematic of the Risk Feature Discovery Pipeline}
%   \label{fig:pipeline}
% \end{figure}
We consider two complexity regimes: \textit{(i) Single-Page Regime ($d=24$)}, capturing lower-order interactions (e.g., local layout shifts, font noise), and \textit{(ii) Multi-Page Regime ($d=27$)}, introducing higher-order dependencies such as cross-page table consistency and split layouts.
The oracle function $f(\mathbf{x})$ returns a scalar risk score computed as the aggregate error of the IDP pipeline across OCR accuracy, layout analysis, table structure recognition, and key-value extraction correctness, with each evaluation requiring a full document generation and end-to-end pipeline execution.

\begin{table*}[t]
\centering
\small
\caption{Comprehensive performance metrics under fixed budget $B=1000$ (Mean of 3 seeds). \textbf{Max}: Best risk score. \textbf{T10$\mu$/$\sigma$}: Mean/Std.Dev of Top-10\% solutions. \textbf{UL}: Unique high-risk configurations. \textbf{Ex\%}: Exclusive Yield. \textbf{Bold}=Best, \underline{Underline}=Second Best (among solvers). Random sampling is given for baseline reference and is not considered for results analysis.}
\setlength{\tabcolsep}{5pt}
\renewcommand{\arraystretch}{1}
\begin{tabular}{ll|cccccc|cccccc}
\hline
& & \multicolumn{6}{c|}{\textbf{Single-Page Regime (24D)}} & \multicolumn{6}{c}{\textbf{Multi-Page Regime (27D)}} \\
\textbf{Class} & \textbf{Method} 
& \textbf{Max} & \textbf{Mean} & \textbf{T10$\mu$} & \textbf{T10$\sigma$} & \textbf{UL} & \textbf{Ex\%} 
& \textbf{Max} & \textbf{Mean} & \textbf{T10$\mu$} & \textbf{T10$\sigma$} & \textbf{UL} & \textbf{Ex\%} \\
\hline
\multirow{2}{0pt}{Local}
& Random & 4.09 & 2.33 & 3.49 & 0.23 & 860 & 10.9 & 4.68 & 2.36 & 3.65 & 0.28 & 981 & 12.1 \\
& SA & 3.98 & 2.97 & 3.96 & 0.02 & 209 & 0.7 & 4.65 & 3.50 & 4.47 & 0.04 & 292 & 1.2 \\
\hline
\multirow{3}{0pt}{Evol}
& GA-Explore & 4.10 & 3.25 & \textbf{4.10} & 0.05 & 378 & 3.2 & 5.06 & 3.81 & \textbf{4.54} & 0.12 & 325 & 2.1 \\
& GA-Exploit & 4.13 & \textbf{3.73} & 4.07 & 0.01 & 109 & 0.4 & 4.50 & \textbf{4.25} & 4.50 & 0.00 & 102 & 0.5 \\
& PSO & 3.92 & 2.32 & 3.61 & 0.12 & 341 & 2.6 & \textbf{5.75} & 2.55 & 4.10 & \underline{0.39} & 614 & \underline{7.7} \\
\hline
\multirow{1}{0pt}{QD}
& MAP-Elites & 4.14 & 2.46 & 3.56 & 0.18 & 466 & 5.1 & 4.71 & 2.69 & 4.07 & 0.19 & 671 & \textbf{9.0} \\
\hline
\multirow{4}{0pt}{Bayes}
& GP-EI & 4.23 & 3.07 & 3.87 & 0.12 & \textbf{957} & \textbf{7.7} & 5.32 & 3.41 & 4.33 & 0.28 & \textbf{994} & 3.3 \\
& GP-UCB & 4.18 & 3.11 & 3.85 & 0.09 & \textbf{957} & \underline{7.6} & 5.15 & 3.43 & 4.33 & 0.22 & \underline{993} & 3.3 \\
& TPE & \underline{4.30} & \underline{3.47} & 3.96 & 0.03 & 187 & 1.2 & 4.44 & 3.94 & 4.44 & 0.00 & 242 & 2.1 \\
\hline
\multirow{3}{0pt}{RL}
& REINFORCE & 4.18 & 2.33 & 3.71 & 0.13 & 809 & 7.5 & 5.16 & 2.81 & 4.29 & 0.18 & 850 & 6.8 \\
& PPO-Div & \textbf{4.40} & 2.87 & 3.85 & 0.15 & 687 & 4.7 & 4.92 & 3.21 & 4.41 & 0.16 & 724 & 5.0 \\
& PPO-Risk & 4.30 & 3.04 & 3.84 & 0.11 & 568 & 3.5 & 5.30 & 3.48 & \underline{4.52} & 0.15 & 560 & 3.3 \\
\hline
\multirow{4}{0pt}{QAOA}
& FM-QAOA & 4.13 & 2.66 & 3.84 & 0.11 & 651 & 3.6 & 4.91 & 3.12 & 4.44 & 0.15 & 701 & 4.6 \\
& FM-QAOA-Corr & 3.99 & 2.37 & 3.73 & 0.12 & 744 & 5.1 & 4.76 & 2.82 & 4.22 & 0.19 & 761 & 5.4 \\
& \textbf{QET-QAOA} & 4.29 & 2.05 & 3.16 & \textbf{0.37} & 462 & 3.8 & 4.67 & 2.35 & 3.89 & 0.34 & 679 & 4.9 \\
& \textbf{QET-QAOA-Corr} & 4.10 & 2.01 & 3.18 & \textbf{0.37} & 520 & 4.4 & \underline{5.54} & 2.42 & 3.85 & \textbf{0.39} & 711 & 5.3 \\
\hline
\end{tabular}
\label{tab:full_results}
\end{table*}

\subsection{Baseline Algorithms}We compare QET-QAOA against representative solvers from four major optimization families. All methods use standard libraries and identical hyperparameters where applicable.

% \textbf{1. Evolutionary and Swarm (Population-Based):} \textbf{Genetic Algorithms (GA):} We evaluate both \textit{Exploration-biased} (high mutation $p_m=0.1$) and \textit{Exploitation-biased} (high crossover $p_c=0.9$) variants using \texttt{DEAP}. \textbf{Particle Swarm Optimization (PSO):} A binary PSO implementation (\texttt{pyswarms}) with inertia weight $w=0.7$.
% \textbf{2. Quality-Diversity (QD):}\textbf{MAP-Elites:} Implemented using \texttt{ribs}. Maintains a $25 \times 25$ feature grid to enforce diversity. This serves as the primary baseline for "coverage" of the search space.
% \textbf{3. Bayesian Optimization (Surrogate-Based):}\textbf{GP-EI / GP-UCB:} Gaussian Process surrogates using \texttt{scikit-learn} with EI and UCB acquisition. \textbf{TPE:} Tree-structured Parzen Estimator implemented in \texttt{Optuna}, optimized for discrete spaces.
% \textbf{4. Reinforcement Learning (Policy Gradient):}Implemented using \texttt{Stable-Baselines3}. \textbf{PPO / REINFORCE:} On-policy gradient methods trained to maximize risk reward. \textbf{PPO-Div} includes an entropy regularization term (coeff $= 0.03$) to encourage exploration.

\textbf{Evolutionary and Swarm (Population-Based):} {Genetic Algorithms (GA)} implemented in \texttt{DEAP}, including an exploration-biased variant (high mutation, $p_m=0.1$) and an exploitation-biased variant (high crossover, $p_c=0.9$); and {Particle Swarm Optimization (PSO)} using a binary PSO implementation (\texttt{pyswarms}) with inertia weight $w=0.7$.

\textbf{Quality-Diversity (QD):} {MAP-Elites} implemented using \texttt{ribs}, maintaining a $25 \times 25$ feature grid to enforce diversity and serve as the primary coverage-oriented baseline.

\textbf{Bayesian Optimization (Surrogate-Based):} {GP-EI} and {GP-UCB} using Gaussian Process surrogates from \texttt{scikit-learn}, and {TPE} (Tree-structured Parzen Estimator) implemented in \texttt{Optuna} for discrete search spaces.

\textbf{Reinforcement Learning (Policy Gradient):} {PPO} and {REINFORCE} implemented in \texttt{Stable-Baselines3}, trained to maximize risk reward; {PPO-Div} additionally incorporates entropy regularization (coefficient $=0.03$) to promote exploration.

\subsection{Quantum Implementation} QET-QAOA is implemented as follows: \textit{Surrogate Model (QET):} A Latent-Quadratic Interaction Embedding Transformer in \texttt{torch} with 2 self-attention layers (4 heads, hidden dim=16 each) is trained via Adam with learning rate of 0.001 with weight decay of 1e-4 for 500 steps. We compare this against FM as surrogates.
\textit{Circuit Ansatz:} We utilize a QAOA circuit of depth $p=2$, using the \texttt{PennyLane} framework with the \texttt{lightning.kokkos} backend. Variational parameters $(\boldsymbol{\gamma}, \boldsymbol{\beta})$ are optimized via COBYLA for 500 iterations per sampling step. We evaluate two mixing Hamiltonians: \textit{(i) Standard Mixer (QAOA):} $H_M = \sum_i X_i$ (Induces independent bit flips), \textit{(ii) Correlated Mixer (QAOA-Corr):} $H_M = \sum_i X_i + \lambda \sum_{(i,j) \in \mathcal{E}} X_i X_j$, utilizing the top-k strong interactions from $Q$ to induce structured transitions.

% \subsection{Evaluation Metrics}To rigorously quantify "Discovery," we move beyond simple scalar max-risk.
% \begin{itemize}\item \textbf{Exclusive Yield (\%):} The percentage of valid high-risk configurations found \emph{only} by a specific solver and missed by all others.\item \textbf{Set Coverage ($|S|$):} The total count of unique, high-utility configurations (risk $> \tau$) discovered.\item \textbf{Jaccard Similarity:} Measures the overlap between the solution sets of two solvers. Low overlap indicates complementary exploration.\end{itemize}
\subsection{Implementation Details} Each method is evaluated with a limited budget of $B=1000$ oracle calls to simulate a constrained evaluation environment. All surrogate and learning-based methods use the same random initial design comprising $N_{\text{init}}=100$ configurations. At each iteration, $50$ samples are selected or generated by each method. The average scores across 3 seeds are reported in quantitative results. 
% Each method executed as an isolated process with a maximum allocation of 64\,GB RAM and 16 CPU cores.

\subsection{Comparative Efficacy} Table~\ref{tab:full_results} illustrates the performance of Quantum-assisted solvers relative to classical baselines across two complexity regimes.

\subsubsection{Exploitation vs Exploration:} We compare how good each method is in finding not only high risks (optimize) but also diversify (sample), by comparing across the max, mean, std.dev. risk scores. In the 24D regime, classical baselines like PPO-Div (4.40), TPE (4.30), and GP-EI (4.23) excel at climbing local gradients. However, their success is brittle. The T10$\sigma$ (Std. Dev of Top-10\%) metric reveals a critical flaw: GA-Exploit, SA, and TPE all show near-zero variance ($\le 0.05$). This indicates they converge to a \textit{single} failure mode and repeatedly resample it. In contrast, Quantum methods like QET-QAOA (4.29) exhibits both competitive max score and also variance; QET-QAOA-Corr maintains a T10$\sigma$ of 0.37 (24D) and 0.39 (27D). Even while achieving the second-highest overall Max Risk (5.54), the $\sigma$ is relatively higher. This shows the quantum sampler is effectively exploring not just the single most top value, but multiple distinct basins simultaneously, a property unique to quantum superposition.
\subsubsection{Topological Exclusivity versus all other methods:} To understand the unique exploratory capabilities of each search strategy, we perform a pairwise analysis of the core risk modes discovered by each method at the end of its budget. We consider the discovery aspect through set-theoritic exclusivity, as described in Section \ref{set_theoric_exclusivity}. 
% In 24D, multiple classical methods do reasonable well in terms of unique layout discovery and good exclusivity, e.g. GP (7.7\%), and REINFORCE (7.5\%), which makes Quantum methods score less.
In the 24D setting, several classical methods already achieve substantial unique layout discovery and high exclusivity (e.g., GP at 7.7\% and REINFORCE at 7.5\%), which narrows the relative exclusivity gains of quantum-based methods.
However, in 27D its low exclusivity (3.3\%) suggests it might be harder to identify newer configurations with increase in dimensionality.
% True novelty lies in the \textbf{Exclusive Yield (Ex\%)}.
MAP-Elites (9.0\%) and PSO (7.7\%) lead in exclusivity in 27D, suggesting simpler diversity-driven objectives might have better reach in high dimensions with constrained budgets. QAOA-Corr methods (FM 5.4\% and QET 5.3\%) remain competitive in this tier, outperforming all other evolutionary and Bayesian methods. This indicates that the correlated mixer is able to sample the search subspace relatively better than gradient-based manifolds in higher dimensions.

\subsubsection{Pairwise Exclusivity of Core Risk Structures:} \label{sec:pairwise_analysis}
To differentiate amongst minor configuration variants, we define \emph{core risk modes} by collapsing failures to their essential semantic predicates (failure type, density regime, noise regime, and layout interaction), creating a set of fundamental structural properties 
Tables~\ref{tab:exclusivity_24d} and~\ref{tab:exclusivity_27d} report pairwise exclusivity of \emph{core risk modes} in the 24D and 27D regimes, respectively, quantifying shared and solver-specific discoveries across methods.

\textbf{24D (Single-Page Regime).} In the lower-dimensional setting, quantum-guided methods consistently outperform classical baselines in exclusive core mode discovery. In particular, QET-QAOA with a correlated mixer (QET-C) acts as a near-superset of most classical solvers, discovering multiple core modes missed by BO methods (e.g., GP-EI, TPE), Simulated Annealing, and MAP-Elites, while rarely missing modes found by them. Among quantum variants, both the QET surrogate and correlated mixing systematically improve exclusivity, indicating that modeling higher-order interactions enables traversal across energy barriers that trap local or purely stochastic search strategies.

\textbf{27D (Multi-Page Regime).} In the higher-dimensional setting, MAP-Elites achieves the strongest exclusivity across most pairwise comparisons, consistent with the scalability of archive-based illumination under increasing combinatorial complexity. Nevertheless, QET-QAOA remains competitive, often ranking second and consistently outperforming BO and RL baselines. The advantage of correlated quantum mixers diminishes in this regime, suggesting that reliable estimation of higher-order correlations requires more data as dimensionality increases. 

Overall, the pairwise analysis confirms that no single solver dominates across regimes: MAP-Elites excels in high-dimensional illumination, while quantum-guided methods provide complementary discovery capabilities by uncovering exclusive core risk modes missed by classical optimizers.

\begin{table}[t!]
    \centering
    \small
    \setlength{\tabcolsep}{1.5pt}
    \renewcommand{\arraystretch}{1.1}
    \caption{
        % \textbf{Pairwise Exclusivity.}
        Comparing pairwise mode exclusivity for 24D (Single Page).
        \textbf{Format:} $\mathbf{Shared} \: / \: \mathbf{Row}_{\text{Excl}} \: / \: \mathbf{Col}_{\text{Excl}}$.
        To read col-wise: Green if Col $>$ Row, Red if Row $>$ Col, Orange if Col $=$ Row.
        % \textbf{Key Result:} \textbf{FM-Corr} achieves high overlap while maintaining exclusivity, acting as a superset to baselines like SA and GP-EI.
    }
    \label{tab:exclusivity_24d}
    % THE FIX: Define 9 columns (1 for labels + 8 for data)
    \begin{tabular}{l|cccccccc}
        \toprule
        \textbf{Method} & \textbf{GP-EI} & \textbf{MAP-E} & \textbf{PPO} & \textbf{QET} & \textbf{FM} & \textbf{QET-C} & \textbf{FM-C} & \textbf{SA} \\
        \midrule
        \textbf{GP-EI}  & \cellcolor{gray!20}-- & \gradcell{18}{6}{1}{8} & \gradcell{21}{3}{3}{8} & \gradcell{23}{1}{1}{8} & \gradcell{24}{0}{1}{8} & \gradcell{24}{0}{3}{8} & \gradcell{24}{0}{2}{8} & \gradcell{21}{3}{0}{8} \\
        \textbf{MAP-E}  & \gradcell{18}{1}{6}{8} & \cellcolor{gray!20}-- & \gradcell{19}{0}{5}{8} & \gradcell{19}{0}{5}{8} & \gradcell{18}{1}{7}{8} & \gradcell{19}{0}{8}{8} & \gradcell{19}{0}{7}{8} & \gradcell{15}{4}{6}{8} \\
        \textbf{PPO}    & \gradcell{21}{3}{3}{8} & \gradcell{19}{5}{0}{8} & \cellcolor{gray!20}-- & \gradcell{21}{3}{3}{8} & \gradcell{22}{2}{3}{8} & \gradcell{24}{0}{3}{8} & \gradcell{23}{1}{3}{8} & \gradcell{18}{6}{3}{8} \\
        \textbf{QET-QAOA}  & \gradcell{23}{1}{1}{8} & \gradcell{19}{5}{0}{8} & \gradcell{21}{3}{3}{8} & \cellcolor{gray!20}-- & \gradcell{23}{2}{1}{8} & \gradcell{24}{0}{3}{8} & \gradcell{24}{0}{2}{8} & \gradcell{20}{4}{1}{8} \\
        \textbf{FM-QAOA}   & \gradcell{24}{1}{0}{8} & \gradcell{18}{7}{1}{8} & \gradcell{22}{3}{2}{8} & \gradcell{23}{1}{2}{8} & \cellcolor{gray!20}-- & \gradcell{25}{0}{2}{8} & \gradcell{25}{1}{0}{8} & \gradcell{21}{4}{0}{8} \\
        \textbf{QET-Corr} & \gradcell{24}{3}{0}{8} & \gradcell{19}{8}{0}{8} & \gradcell{24}{3}{0}{8} & \gradcell{24}{3}{0}{8} & \gradcell{25}{2}{0}{8} & \cellcolor{gray!20}-- & \gradcell{26}{1}{0}{8} & \gradcell{21}{6}{0}{8} \\
        \textbf{FM-Corr}  & \gradcell{24}{2}{0}{8} & \gradcell{19}{7}{0}{8} & \gradcell{23}{3}{1}{8} & \gradcell{24}{2}{0}{8} & \gradcell{25}{0}{1}{8} & \gradcell{26}{0}{1}{8} & \cellcolor{gray!20}-- & \gradcell{21}{5}{0}{8} \\
        \textbf{SA}     & \gradcell{21}{0}{3}{8} & \gradcell{15}{6}{4}{8} & \gradcell{18}{3}{6}{8} & \gradcell{20}{1}{4}{8} & \gradcell{21}{0}{4}{8} & \gradcell{21}{0}{6}{8} & \gradcell{21}{0}{5}{8} & \cellcolor{gray!20}-- \\
        \bottomrule
    \end{tabular}
\end{table}
\begin{table}[t!]
    \centering
    \small
    \setlength{\tabcolsep}{1.5pt} % Note: 1.pt is not a standard unit, changed to 1.5pt
    \renewcommand{\arraystretch}{1.1}
    \caption{
        % \textbf{Pairwise Exclusivity.}
        Comparing pairwise mode exclusivity for 27D (Multi Page).
        \textbf{Format:} $\mathbf{Shared} \: / \: \mathbf{Row}_{\text{Excl}} \: / \: \mathbf{Col}_{\text{Excl}}$.
        To read col-wise: Green if Col $>$ Row, Red if Row $>$ Col, Orange if Col $=$ Row.
        % \textbf{Key Result:} \textbf{FM-Corr} (Row 7) maintains strong exclusivity against gradient-free baselines and achieves complementarity with PPO.
    }
    \label{tab:exclusivity_27d}
    % THE FIX: Define 9 columns (1 for labels + 8 for data)
    \begin{tabular}{l|cccccccc}
        \toprule
        \textbf{Method} & \textbf{GP-EI} & \textbf{MAP-E} & \textbf{PPO} & \textbf{QET} & \textbf{FM} & \textbf{QET-C} & \textbf{FM-C} & \textbf{SA} \\
        \midrule
        \textbf{GP-EI}  & \cellcolor{gray!20}-- & \gradcell{37}{3}{7}{14} & \gradcell{33}{7}{4}{14} & \gradcell{36}{4}{5}{14} & \gradcell{36}{4}{3}{14} & \gradcell{32}{8}{4}{14} & \gradcell{37}{3}{2}{14} & \gradcell{33}{7}{0}{14} \\
        \textbf{MAP-E}  & \gradcell{37}{7}{3}{14} & \cellcolor{gray!20}-- & \gradcell{35}{9}{2}{14} & \gradcell{38}{6}{3}{14} & \gradcell{36}{8}{3}{14} & \gradcell{33}{11}{3}{14} & \gradcell{36}{8}{3}{14} & \gradcell{30}{14}{3}{14} \\
        \textbf{PPO}    & \gradcell{33}{4}{7}{14} & \gradcell{35}{2}{9}{14} & \cellcolor{gray!20}-- & \gradcell{35}{2}{6}{14} & \gradcell{35}{2}{4}{14} & \gradcell{30}{7}{6}{14} & \gradcell{34}{3}{5}{14} & \gradcell{29}{8}{4}{14} \\
        \textbf{QET-QAOA}  & \gradcell{36}{5}{4}{14} & \gradcell{38}{3}{6}{14} & \gradcell{35}{6}{2}{14} & \cellcolor{gray!20}-- & \gradcell{37}{4}{2}{14} & \gradcell{35}{6}{1}{14} & \gradcell{36}{5}{3}{14} & \gradcell{33}{8}{0}{14} \\
        \textbf{FM-QAOA}   & \gradcell{36}{3}{4}{14} & \gradcell{36}{3}{8}{14} & \gradcell{35}{4}{2}{14} & \gradcell{37}{2}{4}{14} & \cellcolor{gray!20}-- & \gradcell{34}{5}{2}{14} & \gradcell{37}{2}{2}{14} & \gradcell{33}{6}{0}{14} \\
        \textbf{QET-Corr} & \gradcell{32}{4}{8}{14} & \gradcell{33}{3}{11}{14} & \gradcell{30}{6}{7}{14} & \gradcell{35}{1}{6}{14} & \gradcell{34}{2}{5}{14} & \cellcolor{gray!20}-- & \gradcell{33}{3}{6}{14} & \gradcell{32}{4}{1}{14} \\
        \textbf{FM-Corr}  & \gradcell{37}{2}{3}{14} & \gradcell{36}{3}{8}{14} & \gradcell{34}{5}{3}{14} & \gradcell{36}{3}{5}{14} & \gradcell{37}{2}{2}{14} & \gradcell{33}{6}{3}{14} & \cellcolor{gray!20}-- & \gradcell{33}{6}{0}{14} \\
        \textbf{SA}     & \gradcell{33}{0}{7}{14} & \gradcell{30}{3}{14}{14} & \gradcell{29}{4}{8}{14} & \gradcell{33}{0}{8}{14} & \gradcell{33}{0}{6}{14} & \gradcell{32}{1}{4}{14} & \gradcell{33}{0}{6}{14} & \cellcolor{gray!20}-- \\
        \bottomrule
    \end{tabular}
\end{table}

\begin{figure*}[t]
  \centering
  \includegraphics[width=0.95\linewidth]{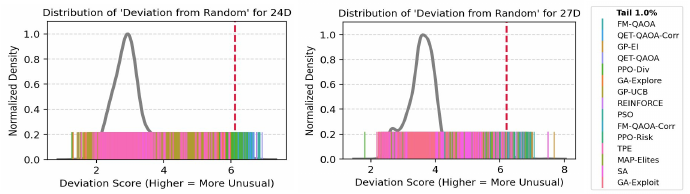}
\caption{Tail distribution plot across all methods, for 24D and 27D regimes. 
}
  \label{fig:tail}
\end{figure*}

\subsubsection{Distributional Tail Sampling:}
While exclusive discovery captures \emph{what} failures are found, effective stress testing also requires measuring \emph{tail diversity}: the ability to discover structurally extreme configurations far from typical solutions. We quantify structural atypicality using a deviation score, defined as the standardized Euclidean distance from a random-sampled centroid in configuration space. Table~\ref{tab:outlier_discovery} reports the number of configurations falling within the top 1\% of this deviation distribution, while Figure~\ref{fig:tail} shows the full deviation-score distributions.

\begin{table}[h]
\centering
\caption{\textbf{Distributional Tail Sampling.} Number of configurations discovered within the 1\% structural deviation tail, out of 1000 samples per method.}
\label{tab:outlier_discovery}
\setlength{\tabcolsep}{4pt}
\begin{tabular}{l | c | c}
\toprule
\textbf{Method} & \textbf{24D Tail (1\%)} & \textbf{27D Tail (1\%)} \\
\midrule
\textbf{FM-QAOA-Corr}   & \textbf{27} & 18 \\
\textbf{FM-QAOA}        & 12 & \textbf{26} \\
\textbf{QET-QAOA-Corr}  & \underline{23} & \underline{21} \\
\textbf{QET-QAOA}       & 20 & 14 \\
MAP-Elites              & 18 & 2  \\
REINFORCE               & 16 & 13 \\
PPO-Risk                & 12 & 7  \\
GP-EI                   & 0  & 12 \\
GP-UCB                  & 0  & 10 \\
GA-Explore              & 0  & 10 \\
TPE                     & 0  & 3  \\
SA                      & 0  & 2  \\
GA-Exploit              & 0  & 1  \\
\bottomrule
\end{tabular}
\vspace{0.1cm}
\end{table}
Two clear patterns emerge. {First}, several objective-driven optimizers with strong average performance contribute few or no tail elites in the 24D regime, indicating concentration around dominant risk structures. While diversity-oriented MAP-Elites improves tail coverage, it remains below the best quantum methods. {Second}, QAOA-based samplers consistently rank at or near the top in tail discovery across both dimensions, for example, FM-QAOA-Corr in 24D and FM-QAOA in 27D, with QET-QAOA-Corr remaining consistently competitive, demonstrating more reliable recovery of structurally extreme risk configurations.
Figure~\ref{fig:tail} illustrates the underlying mechanism. Classical optimizers concentrate sampling mass near central, frequently visited structures, whereas surrogate-guided QAOA maintains heavier distributional tails. Because QAOA samples from a learned energy-shaped distribution rather than iteratively collapsing toward a single improving mode, multiple high-risk basins,including structurally distant ones, retain non-negligible probability mass under fixed evaluation budgets.

\subsubsection{Surrogate Generalizability \& Information Gain}
\label{sec:surrogate_modeling}
Beyond immediate risk discovery, a critical property of a search strategy is the information value of the samples it generates. Does the solver explore the landscape informatively, producing data that allows a model to generalize to unseen regions? This reframes the problem to asking which solver generates the best training data for mapping the global risk function?
To quantify this, we treat the sample history of each solver ($N=700$) as a training set for a Random Forest Regressor and evaluate its predictive accuracy ($R^2$) on the remaining 30\% from each method to form the collective global hold-out validation set.

The results in Table~\ref{tab:surrogate_performance} demonstrate the superior information gain of the quantum-guided search: \textbf{Quality over Quantity:} In the 24D regime, the model trained on data from QET-QAOA-Corr achieves the highest generalization score ($R^2=0.840$), significantly outperforming models trained on RL (0.78\%), GP-EI ($R^2=0.652$) or TPE ($R^2=0.658$). This implies that the quantum sampler covers the landscape's geometry more comprehensively compared to other methods. \textbf{Robustness:} While GP-EI performs well in 27D ($R^2=0.757$), FM-QAOA ($0.741$) and QET-QAOA ($0.740$) remain statistically equivalent. While Gaussian Process (GPs) suffer from cubic computational complexity ($O(N^3)$) \cite{rasmussen2006gaussian} with respect to the number of samples,
% our parametric QET surrogate scales linearly ($O(N)$), 
our parametric QET transformer surrogate reduces this to quadratic complexity (\(O(N^{2})\)) (with scope to be further brought down to (\(O(N)\)) using linear attention \cite{katharopoulos2020transformers}), enabling efficient learning in high-dimensional spaces where traditional Bayesian Optimization becomes computationally prohibitive.
\begin{table}[t!]\centering
\small
\caption{\textbf{Surrogate Generalizability ($R^2$).} Models are trained \textit{only} on 70\% of samples discovered by each specific method, and tested on the hold-out 30\% validation. A higher $R^2$ indicates the method explored the landscape more informatively, finding samples across regions leading to generalization.}
\vspace{-0.15cm}
\label{tab:surrogate_performance}
\setlength{\tabcolsep}{3.5pt}\begin{tabular}{l | ccc | ccc}
\toprule & \multicolumn{3}{c|}{\textbf{24D}} & \multicolumn{3}{c}{\textbf{27D}} \\
\cmidrule(lr){2-4} 
\cmidrule(lr){5-7}
\textbf{Training Source} & $\mathbf{R^2 \uparrow}$ & \textbf{MAE $\downarrow$} & \textbf{MAPE} & $\mathbf{R^2 \uparrow}$ & \textbf{MAE $\downarrow$} & \textbf{MAPE} \\
\midrule
\textit{Reference (All Data)} & \textit{0.924} & \textit{0.157} & \textit{0.068} & \textit{0.924} & \textit{0.174} & \textit{0.064} \\
\midrule
\textbf{QET-QAOA-Corr}      & \textbf{0.840} & \textbf{0.269} & \textbf{0.107} & 0.729 & 0.374 & 0.152 \\
\textbf{FM-QAOA-Corr}       & \underline{0.824} & \underline{0.279} & \underline{0.123} & 0.726 & 0.369 & 0.147 \\
\textbf{QET-QAOA}           & 0.798 & 0.293 & 0.120 & 0.740 & 0.388 & \textbf{0.135} \\
\textbf{FM-QAOA}            & 0.785 & 0.299 & 0.140 & \underline{0.741} & \underline{0.386} & 0.146 \\
\midrule
REINFORCE                   & 0.787 & 0.297 & 0.117 & 0.719 & 0.365 & \textbf{0.123} \\
PPO-Div                     & 0.781 & 0.305 & 0.146 & 0.712 & 0.378 & 0.131 \\
GP-EI                       & 0.652 & 0.364 & 0.216 & \textbf{0.757} & \textbf{0.350} & 0.158 \\
MAP-Elites                  & 0.681 & 0.376 & 0.193 & 0.573 & 0.478 & 0.166 \\
TPE                         & 0.658 & 0.376 & 0.211 & 0.652 & 0.441 & 0.175 \\
Random                      & 0.679 & 0.332 & 0.197 & 0.510 & 0.502 & 0.179 \\
\bottomrule
\end{tabular}\end{table}

\subsection{Architectural Ablation (QET vs.\ FM)}
\label{sec:ablation}
We compare the Latent-Quadratic Embedding Transformer (QET) and Factorization Machine (FM) surrogates under identical data, training budgets, and QAOA settings. Both produce quadratic surrogate Hamiltonians; they differ only in interaction modeling: FM encodes explicit pairwise terms, whereas QET captures higher-order dependencies via contextual self-attention before quadratic projection.

\textbf{Surrogate fidelity:} Predictive accuracy is comparable across regimes (Table~\ref{tab:surrogate_performance}). In 24D, QET generalizes slightly better ($R^2=0.840$ vs.\ $0.824$), while in 27D both achieve similar performance ($R^2 \approx 0.73--0.74$), indicating that downstream differences are not driven by prediction error.

\textbf{Induced search behavior:} Despite similar fidelity, the induced QAOA sampling distributions differ. In 24D, where interactions are mostly pairwise, FM provides broader coverage and higher mean sampled risk (UL: 744 vs.\ 462), while QET reaches higher-risk extremes (max: 4.29 vs.\ 4.13/3.99). In 27D, FM maintains larger coverage (UL: 761 vs.\ 272), whereas QET accesses substantially higher extremes (max: 5.54 vs.\ 4.76), consistent with sensitivity to higher-order interactions.

\textbf{Elite diversity:} QET consistently produces more structurally diverse elites (Top-10\% variance: 0.37/0.39 vs.\ FM 0.11/0.19), indicating that contextual interaction modeling reshapes the quadratic energy landscape so that multiple high-risk basins remain competitive.
Overall, FM favors broad coverage through pairwise averaging, whereas QET favors selective extreme discovery and elite diversity, with advantages increasing as interaction complexity grows.

% \vspace{-0.1cm}
\section{Discussion}
\label{sec:discussion}
We study diversity-oriented discovery under strict query budgets, where the objective is to recover a diverse set of high-utility configurations rather than a single optimum. In this black-box regime, surrogate-guided quantum sampling is effective: learning a quadratic proxy and sampling with QAOA shifts search behavior from mode-seeking optimization toward distributional exploration.
Discovery behavior is primarily governed by mixer topology and circuit depth. Interaction-aware correlated mixers improve connectivity between high-utility regions and increase exclusive discovery in structured settings. For sampling, shallow circuits are preferable, as they preserve probability mass across multiple basins, whereas deeper circuits tend to over-concentrate on a few optima, reducing diversity and increasing cost.
Surrogate structure determines which regions of the landscape are amplified. Pairwise surrogates are effective when interactions are simple, but saturate as dimensionality and interaction complexity grow. Contextual surrogates that model higher-order dependencies before quadratic projection reshape the geometry of the learned energy landscape, altering the induced sampling distribution even when predictive accuracy is comparable.
Compared to classical optimizers, which often collapse to dominant modes, surrogate-guided QAOA maintains heavier distributional tails and improves exclusive discovery. The FM–QET ablation reveals a regime-dependent trade-off: pairwise surrogates favor broad coverage, while contextual surrogates improve tail discovery and diversity when higher-order structure is present.

\section{Conclusion}
\label{sec:conclusion}

% This work introduced a surrogate-guided quantum sampling framework that learns a quadratic proxy Hamiltonian via a Transformer-based Latent-Quadratic Interaction embedding model and uses shallow-depth QAOA for diversity-oriented discovery under strict evaluation budgets. The approach improves tail discovery, elite diversity, and exclusivity relative to classical baselines while remaining hardware-efficient. Beyond diversity sampling, QET-style contextual surrogate construction can be repurposed for pure optimization by acting as a learned QUBO feature map that lifts structured inputs into interaction-aware quadratic forms compatible with both quantum and classical solvers. This enables reuse in surrogate-assisted combinatorial optimization and black-box search. Promising next steps include transferring learned quadratic embeddings across related tasks, integrating the surrogate as a plug-in energy model inside classical optimizers, and extending the framework to constrained and multi-objective optimization settings.
This work presents a surrogate-guided quantum sampling framework that learns interaction-aware quadratic proxy Hamiltonians from data by projecting higher-order dependencies, captured via a Transformer-based Latent-Quadratic Interaction embedding, into QAOA-compatible forms. Using shallow-depth QAOA, the framework enables diversity-oriented discovery under strict evaluation budgets, improving tail discovery, elite diversity, and exclusivity relative to classical baselines while remaining hardware-efficient.
Beyond quantum sampling, QET-style contextual surrogate construction functions as a learned QUBO feature map that lifts structured inputs into interaction-aware quadratic representations, making them compatible with both quantum algorithms and classical optimizers. This enables reuse in surrogate-assisted combinatorial optimization and black-box search.
Promising directions include transferring learned embeddings across related tasks, and integrating the surrogate as a plug-in to extend the framework to constrained and multi-objective optimization settings.

\section*{Acknowledgments}
This work is supported by PhiLabs, Quantiphi. We would like to thank Varun V for suggesting editorial improvements to the manuscript, Harkrishnan PM for inputs on the IDP pipeline, and our cofounder Asif Hasan for the continued support.

\clearpage

% \newpage
% \section{ACM's Authorship Policy on Generative AI}
% We acknowledge the use of Google Scholar Labs for literature retrieval, Claude Sonnet 4 for experimental code assistance, and Gemini 3 Pro and GPT 5.2 for editorial refinement. We have verified the accuracy of all AI-generated outputs and retain full responsibility for the scientific content and claims presented herein.


\bibliographystyle{ACM-Reference-Format}
\bibliography{main}

\end{document}